\documentclass[aps,prl,onecolumn,superscriptaddress,floatfix,12pt]{revtex4-1}
\usepackage{amssymb,amsmath}
\usepackage{graphicx}
\usepackage{float}
\usepackage{bm}
\usepackage{multirow}
\usepackage{dcolumn}
\usepackage{tabularx}
\usepackage{longtable}
\usepackage[dvipsnames,usenames]{color}
\usepackage[latin1]{inputenc}

\newlength{\dbarheight}

\begin{document}

\title{Reconciling the ionic and covalent pictures in rare-earth nickelates}

\author{Julien Varignon} \affiliation{Unit\'e Mixte de Physique, CNRS,
  Thales, Universit\'e Paris Sud, Universit\'e Paris-Saclay, 1 avenue A. Fresnel, 91767,
  Palaiseau, France} \email{julien.varignon@thalesgroup.com}

\author{Mathieu N. Grisolia} \affiliation{Unit\'e Mixte de Physique,
  CNRS, Thales, Universit\'e Paris Sud, Universit\'e Paris-Saclay,
  1 avenue A. Fresnel, 91767, Palaiseau, France}

\author{Jorge \'I\~{n}iguez} \affiliation{Materials Research and
  Technology Department, Luxembourg Institute of Science and
  Technology (LIST), 5 avenue des Hauts-Fourneaux, L-4362
  Esch/Alzette, Luxembourg}

\author{Agn\`es Barth\'el\'emy} \affiliation{Unit\'e Mixte de
  Physique, CNRS, Thales, Universit\'e Paris Sud, Universit\'e
  Paris-Saclay, 1 avenue A. Fresnel, 91767, Palaiseau, France}

\author{Manuel Bibes} \affiliation{Unit\'e Mixte de Physique, CNRS,
  Thales, Universit\'e Paris Sud, Universit\'e Paris-Saclay, 1 avenue
  A. Fresnel, 91767, Palaiseau, France}

\date{\today}

\begin{abstract}
The properties of AMO$_3$ perovskite oxides, where M is a $3d$
transition metal, depend strongly on the level of covalency between
the metal $d$ and oxygen $p$ orbitals. With their complex spin orders
and metal-insulator transition, rare-earth nickelates verge between
dominantly ionic and covalent characters. Accordingly, the nature of
their ground state is highly debated. Here, we reconcile the ionic and
covalent visions of the insulating state of nickelates. Through
first-principles calculations, we show that it is reminiscent of the
ionic charge disproportionation picture (with strictly low-spin 4+ and
high-spin 2+ Ni sites) while exhibiting strong covalence effects with
oxygen electrons shifted toward the depleted Ni cations, mimicking a
configuration with identical Ni sites. Our results further hint at
strategies to control electronic and magnetic phases of transition
metal oxide perovskites.
\end{abstract} 
\maketitle

Transition metal oxides with an {\sl AM}O$_3$ perovskite structure
have attracted widespread interest over the last decades, both from
academic and industrial points of view.  This can be ascribed to their
wide range of functionalities that originates from the interplay
between lattice, electronic, and magnetic degrees of
freedom~\cite{zubko2011interface}. Among all perovskites, rare-earth
nickelates {\sl R}$^{3+}$Ni$^{3+}$O$_3$ ({\sl R}=Lu-La, Y) might be
considered as a prototypical case because they posses almost all
possible degrees of freedom present in these materials. Nickelates
were intensively studied during the
nineties~\cite{Review-medarde1997structural,Review-catalan2008progress}
and have regained interest in the few last years due to their great
potential for engineering novel electronic and magnetic
states~\cite{RNO-dimensionality,Giovannetti-RNO,Jorge-RLaNiMnO6,LaNiMO6-HighTc,NPhys-Mathieu,NMat-Marta,Keimer-Reflecto,Keimer-PNOPAO}.
 
Except for {\sl R}=La, all rare-earth nickelates undergo a
metal-insulator phase transition (MIT) at a temperature
T$_{\text{MI}}$, accompanied by a symmetry lowering from $Pbnm$ to
$P2_1/n$~\cite{Review-medarde1997structural,Review-catalan2008progress}. In
this $P2_1/n$ phase, a Ni-site splitting is observed; this is usually
associated with the appearance of charge
disproportionation~\cite{alonso1999metal,PRB-Medarde-PNO,PRB-medarde-2009}
from 2Ni$^{3+}$ to Ni$^{(3+\delta)+}$ + Ni$^{(3-\delta)+}$ and/or a
breathing distortion of O$_{6}$ octahedra that leads to a
rock-salt-like pattern of small and large NiO$_6$
groups~\cite{PRB-Medarde-PNO}. At T$_{\text N}$ $\le$ T$_{\text{MI}}$,
nickelates undergo an antiferromagnetic (AFM) phase transition
yielding a quadrupling of the magnetic unit cell ($\vec
k$=($\frac{1}{2}, 0, \frac{1}{2}$) with respect to the $Pbnm$
primitive cell) and possible collinear or non collinear spin
orderings~\cite{garcia1994neutronPNONNO,AFMT-SNO-EuNO,SpinDW-NNO-PNO,RNO-noncolinearMag}.
The electronic structure is also characterized by strong overlaps
between O-$2p$ and Ni-$3d$ states leading to large covalent
effects~\cite{Review-medarde1997structural}. As a consequence,
external stimuli, such as temperature, or chemical or hydrostatic
pressure, can modify the electronic bandwidth and influence the
MIT~\cite{RNO-bandwidth,RNO-pressure,RNO-pressure2,zhou2005pressure}. Efforts
have thus been devoted to search for novel electronic phases in
nickelates, mainly using strain engineering or
confinement~\cite{scherwitzl2010electric,straincontrol-RNO,SNO-HSE-Lezaic,Keimer-Reflecto,Keimer-PNOPAO}.

In spite of all these research efforts, the structural, electronic and
magnetic properties of the bulk ground state are still under
debate. This can be ascribed to the scarcity of systematic bulk
studies, from both the experimental and theoretical sides. On one
hand, bulk nickelates are hard to synthesize and mainly thin films
have been studied~\cite{Review-catalan2008progress}. On the other
hand, no theoretical systematic studies have been performed due to the
difficulty of reproducing the {\sl R}NiO$_3$ ground state using
density functional theory (DFT). In the context of DFT-based
calculations, the choice of the Hubbard U correction for Ni-$3d$
levels remains ambiguous; indeed, a great diversity of values, ranging
from very weak to quite strong corrections, have been proposed and
argued for in different
works~\cite{NNO-Prosandeev,yamamoto2002charge,Breathing-alternative,Giovannetti-RNO,AGeorges-lowenergyNickelates,straincontrol-RNO}. Moreover,
the identified ground state is usually
ferromagnetic~\cite{straincontrol-RNO,NNO-Prosandeev,park2012site}, in
contrast to the established antiferromagnetic ordering.

Here we performed a systematic study of various representative
nickelates using the standard DFT+U formalism. We find that a small
Coulombian correction on Ni-$3d$ states is appropriate to reproduce
the key ground state properties of these compounds. We then use this
theory to discuss the electronic ground state of the nickelates,
revealing the co-existence of ionic (Ni electronic states featuring a
complete and strict charge disproportionation) and covalent
(oxygen-$p$ electrons shared with the charge-depleted Ni cations)
features, and providing an unified picture of these materials that is
easy to reconcile with existing (and apparently conflicting) proposals
in the literature. Finally, we unveil a new pathway to control
electronic and magnetic phases in perovskites by tuning the level of
covalency.

\begin{figure}
\begin{center}
\resizebox{8.6cm}{!}{\includegraphics{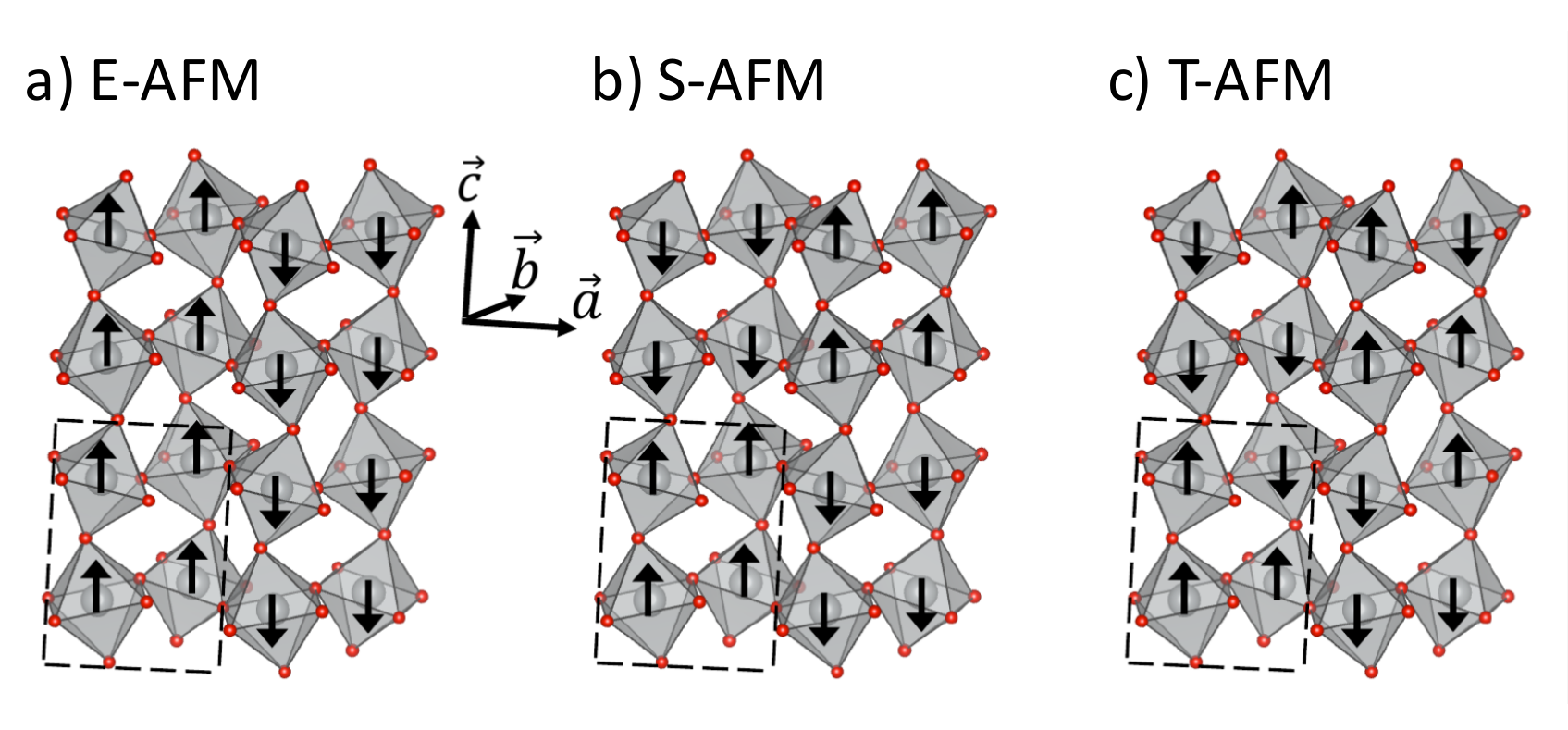}}
\end{center}
\caption{{\bf Sketch of the three complex antiferromagnetic orderings
    used in the calculations.} (a) E-type AFM ordering. (b) S-type AFM
  ordering. (c) T-type AFM ordering.  The A cations are not displayed
  for clarity. The dashed lines represent the size of the
  crystallographic unit cell.}
\label{fig1}
\end{figure}

\section*{Results}
\subsection*{Structural properties}


\begin{figure*}
\begin{center}
\resizebox{17.2cm}{!}{\includegraphics{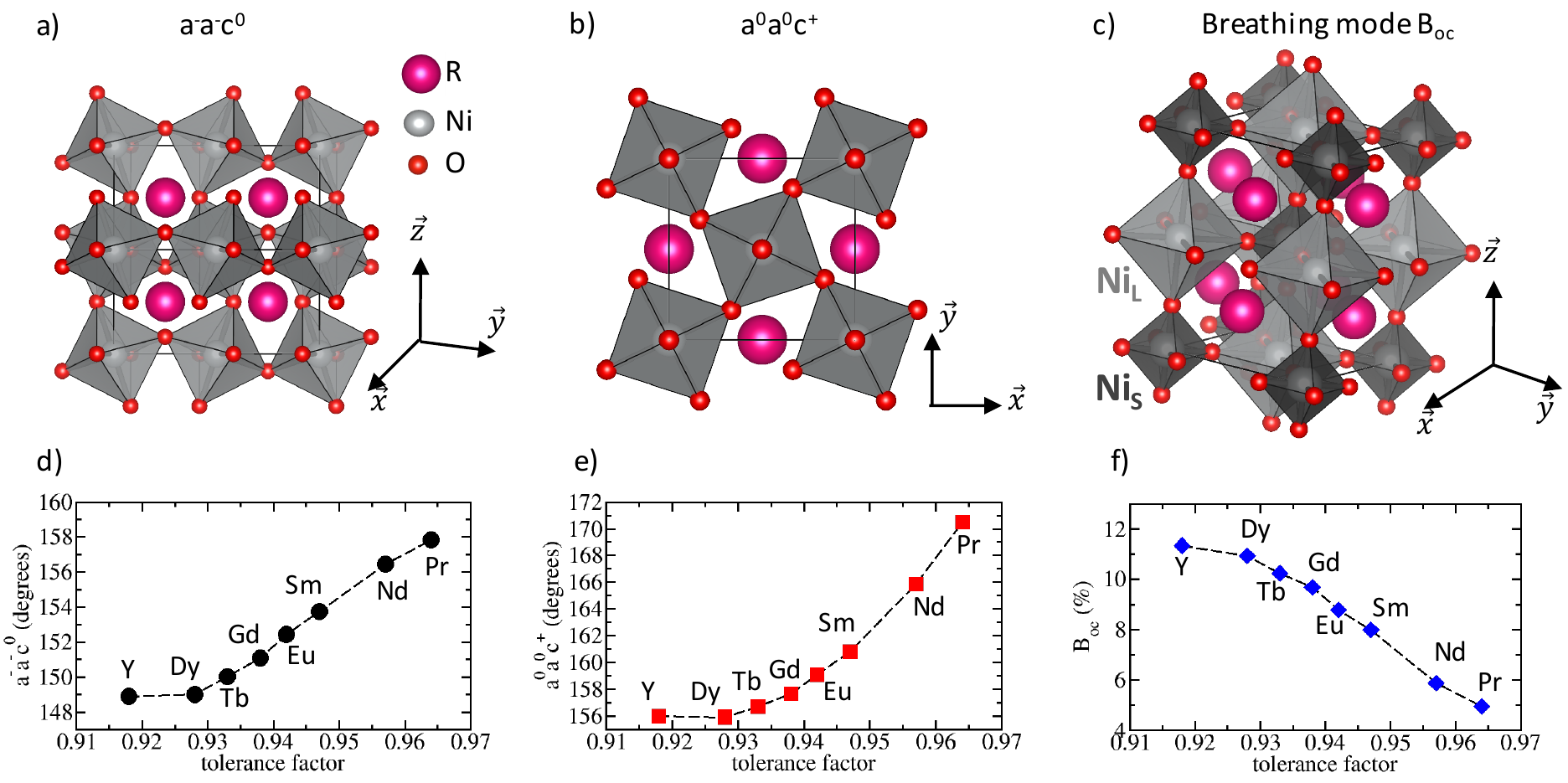}}
\end{center}
\caption{{\bf Rare-earth nickelates ground state structural
    properties.} (a), (b) and (c) Schematic pictures of the three main
  lattice distortions appearing in the ground state of nickelates:
  $a^-a^-c^0$ antiferrodistortive motion, $a^0a^0c^+$
  antiferrodistortive motion and breathing of the oxygen cage
  octahedra B$_{\text{oc}}$.  (d) and (e) Ni-O-Ni angle (in degrees)
  produced by the $a^-a^-c^0$ and $a^0a^0c^+$ antiferrodistortive
  motions. (f) Oxygen cage octahedra volume expansion/contraction (in
  \%) produced by the breathing mode.}
\label{fig2}
\end{figure*}

First, we performed full geometry relaxations considering 80-atom
supercells of both possible $Pbnm$ and $P2_1/n$ structures with
different magnetic orderings: ferromagnetic (FM) as well as complex
E-, S-, and T-type AFM orderings~\cite{Giovannetti-RNO} based on
$\uparrow \uparrow \downarrow \downarrow$ spin chains in the
(ab)-plane with different stackings along the $\vec c$ axis (see
Figures~\ref{fig1}.a, b and c). We employed the PBEsol
functional~\cite{PBEsol} in combination with a {\sl U}
correction~\cite{LDAU-Lich} of 2 eV on Ni-$3d$ states in order to
account for electronic correlations. Several nickelates ({\sl R}=Y,
Dy, Tb, Gd, Eu, Sm, Nd, and Pr) were considered, covering the phase
diagram as a function of rare-earth radius. All nickelates relax to a
$P2_1/n$ insulating ground state with complex antiferromagnetic
structures (S- or T-type depending on the rare-earth) and band gaps
compatible with
experiments~\cite{Tokura-bandgap,BandGap-EvolutionAngleNiONi} (see
Table~\ref{tab1}). All our $Pbnm$ phases favor a metallic FM
solution~\cite{noteOrdreMag}. We checked the reliability of our DFT+U
calculations by changing the U correction to either 0~eV or 5~eV in
SmNiO$_3$. While the ground state is unchanged when no U-correction is
applied~\cite{gap0eV}, imposing U = 5~eV yields a $P2_1/n$
ferromagnetic and insulating solution that is much more stable than
the considered complex AFM orderings ($\Delta E \simeq$ 160 meV per
80-atom unit cell). This further supports our choice of a relatively
small Hubbard correction for the Ni-$3d$ electrons.

\begin{table}
\centering
\begin{tabular}{ccccc}
\hline
R & $\Delta$E(E-AFM) & $\Delta$E(S-AFM) & $\Delta$E(T-AFM) & gap (eV)\\
\hline
Pr & -266 & -393 & -384 & 0.49 \\
Nd & -172 & -290 & -282 & 0.50 \\
Sm & -71  & -139 & -145 & 0.54 \\
Eu & -35  & -82  & -92  & 0.56 \\
Gd & -15  & -51  &  -34 & 0.55 \\
Tb & -9 & -33 & -13 & 0.58\\
Dy & -6  &  -10 & 13 & 0.59\\
Y & -4 &-20 &3 & 0.61\\
\hline
\end{tabular}
\caption{{\bf Key quantities of the different relaxed ground states.}
  Computed energy differences $\Delta$E (in meV per 80-atom unit cell)
  between the complex antiferromagnetic and ferromagnetic solutions
  and electronic band gap of the identified ground state.}
\label{tab1}
\end{table}

\begin{figure}[h!]
\begin{center}
\resizebox{8.6cm}{!}{\includegraphics{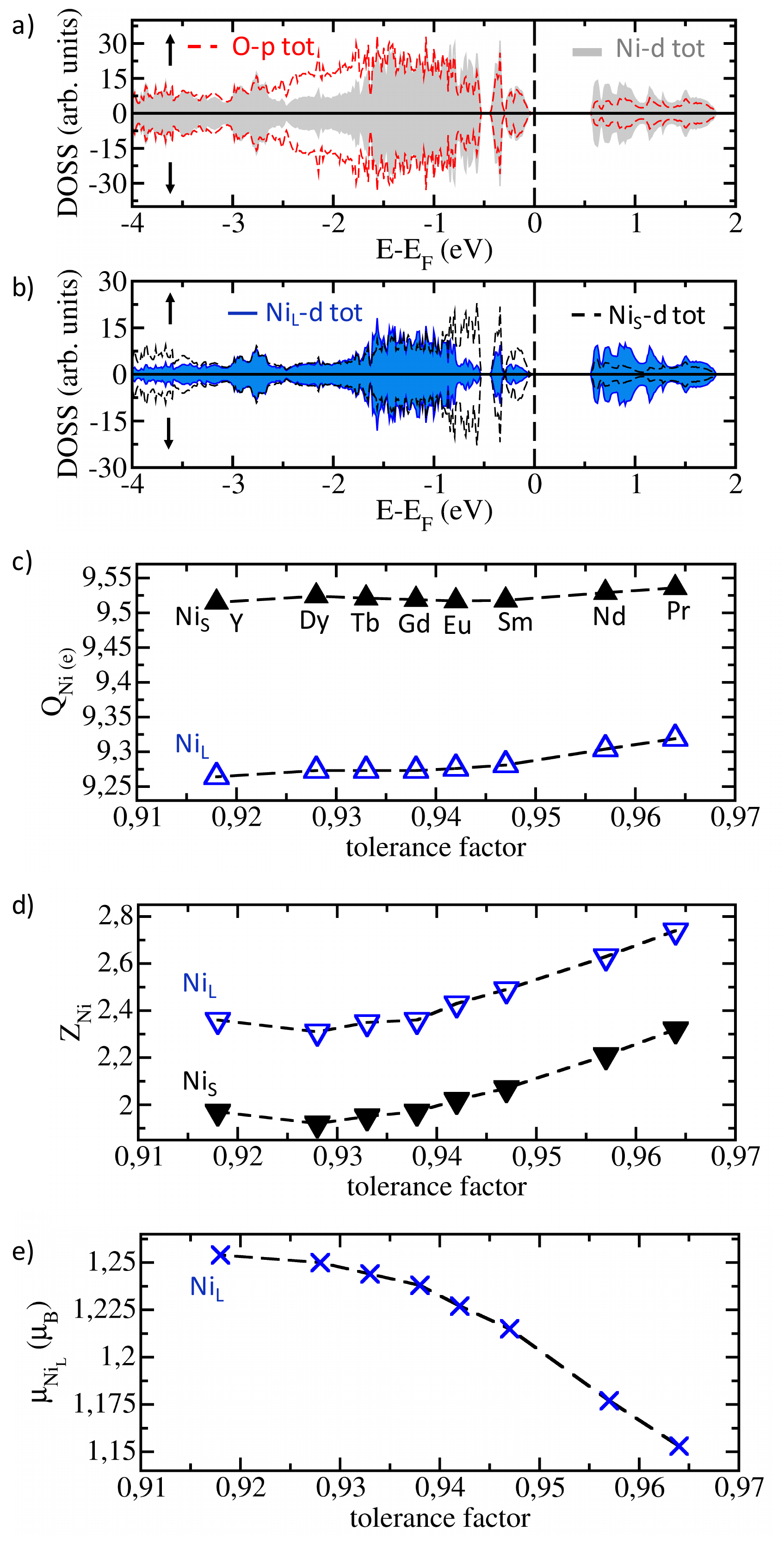}}
\end{center}
\caption{{\bf Key electronic properties of rare-earth nickelates
    ground state.}  (a) and (b) Total and projected spin-polarized
  density of states (arbitrary units). The Fermi level is located at 0
  eV. The upper (lower) panels correspond to spin up (down). (c) Total
  number of electrons ascribed to both Ni sites during the sphere
  integrations.  (d) Average Born effective charges of both Ni sites.
  (e) Computed magnetic moment (in $\mu_B$) of the Ni$_{\rm L}$
  site. The magnetic moment of Ni$_{\rm S}$ is exactly zero in the
  calculation.}
\label{fig3}
\end{figure}

Our optimized ground state structures are characterized by three main
lattice distortions. First, they feature two antiferrodistortive (AFD)
modes that can be described, respectively, as $a^-a^-c^0$ and
$a^0a^0c^+$ patterns using Glazer's notation~\cite{Glazer}. These AFD
modes are the main features of the phase with $Pbnm$ symmetry. Second,
we have a breathing of the O$_{6}$ octahedra, B$_{\text{oc}}$ (see
Figures~\ref{fig2}.a, b and c). The breathing mode only appears in the
$P2_1/n$ symmetry and produces a rock-salt pattern of small and large
NiO$_6$ groups, automatically resulting in two different Ni sites (see
Figure~\ref{fig2}.c). In the following we will use the notation
Ni$_{\rm S}$ and Ni$_{\rm L}$ to refer to the Ni cations belonging to
the small and large NiO$_6$ groups, respectively.

As usual in perovskites, the magnitude of the metal-oxygen-metal bond
angles associated with the O$_{6}$ rotations is governed by steric
effects (see Figures~\ref{fig2}.d and e), and nickelates with low
tolerance factors ({\em i.e.}, smaller {\sl R}
cations)~\cite{goldschmidt1926gesetze} are more distorted. The
alternating expansion/contraction pattern of the oxygen cage
associated with the B${_\text{oc}}$ breathing also appears to be
modulated by the rare earth (see Figure~\ref{fig2}.f), as smaller {\sl
  R} cations yield larger distortions. Finally, we observe a
Jahn-Teller distortion in the ground state that is one to two orders
of magnitude smaller than the breathing mode or the two AFD
motions. Hence, the relaxed structures indicate that there is no
significant orbital order in these systems, although the 3$d^7$
${\text t}_{\text {2g}}^6{\text e}_{\text g}^1$ electronic
configuration of Ni$^{3+}$ in the high temperature $Pbnm$ phase is
nominally Jahn-Teller active~\cite{Breathing-alternative}.


\subsection*{Disproportionation signatures}

The electronic structure of the optimized ground states is
characterized by strong hybridizations between O-$2p$ and Ni-$3d$
levels, as inferred from the projected density of states (pDOS, see
Figure~\ref{fig3}.a for the representative case of
SmNiO$_{3}$). Comparing the pDOS corresponding to the $3d$ levels of
the two different Ni sub-lattices reveals some small differences,
likely reflecting weak disproportionation effects and a small charge
ordering (see Figure~\ref{fig3}.b). Although atomic charges are not
uniquely defined in DFT calculations~\cite{Pickett-chargeDFT}, sphere
integrations around the Ni cations can provide some insight into the
possible charge ordering. Figure~\ref{fig3}.c reports the occupancy of
both Ni sites as a function of the rare earth. A weak and rather
constant charge ordering is observed between the Ni sites, going from
$\delta$=0.13 (YNiO$_3$) to $\delta$=0.11 (PrNiO$_3$), but the sign of
$\delta$ is opposite to what is expected. Indeed, the Ni$_{\rm S}$
cations, sitting at the center of the smallest O$_{6}$ octahedra,
appear to hold more electrons than the Ni$_{\rm L}$ cations, located
in the largest oxygen cages. Since the breathing mode B$_{\text {oc}}$
enhances the crystal field splitting at the small NiO$_6$ groups, the
$e_g$ levels of Ni$_{\rm S}$ lie higher in energy than those of
Ni$_{\rm L}$~\cite{Breathing-alternative,NComms-Titanates} and
therefore Ni$_{\rm S}$ should have fewer electrons than Ni$_{\rm L}$
associated to it.

Let us now consider better defined -- and experimentally measurable --
quantities, such as Born effective charges (BECs) that measure the
amount of charge displaced upon the movement of individual
atoms. Figure~\ref{fig3}.d reports the average of the diagonal
components of the tensor for the different nickelates (see the
Supplementary Material for the full tensors). In the representative
case of SmNiO$_{3}$, we obtain Z$_{\text Ni_{\text L}} \approx +2.5$,
which is not far from the nominal oxidation state of $2+$ that this Ni
site is associated with in the complete-charge-disproportionation
picture. However, we find a similar Z$_{\text Ni_{\text S}} \approx
+2.1$, which sharply deviates from the expectation value ($4+$) in the
charge-disproportionation picture. As shown in Figure~\ref{fig3}.d, we
observe the same behavior across the various studied compounds; an
approximately constant difference of Born charges, of about 0.4
electrons, is found across the whole series. Hence, from the point of
view of the effective charges, the two Ni sites behave in a rather
similar fashion, the disproportionation effects being weak and not
complying with the usual picture.

However, our computed magnetic moments on both Ni sites appear to be
in contradiction with the conclusion of the charge analysis. Indeed,
as shown in Figure~\ref{fig3}.e, we observe a large difference between
Ni$_{\rm L}$ -- with a moment larger than 1~$\mu_{\rm B}$ -- and
Ni$_{\rm S}$ -- for which the magnetic moment is null --, which
suggests two very different electronic states.

\subsection*{Wannier analysis}

The conclusion of the previous discussion is that, in the RNiO$_{3}$
compounds, all Ni atoms seem to display a similar oxidation
state. Yet, the presence of a significant breathing distortion, and of
the drastic difference in the local magnetic moments, clearly suggest
two markedly different electronic states. Note that similar results
have been reported in previous theoretical works using a variety of
methods~\cite{NNO-Prosandeev,SNO-HSE-Lezaic,park2012site}, but in our
opinion a convincing explanation for this apparent contradiction is
still missing. Here we ran a Wannier function (WF) analysis of our
first-principles results, which allowed us to resolve this pending
issue.

We used the Wannier90
package~\cite{Wannier90-1,Wannier90-2,Wannier90-3} to determine the
maximally-localized WFs that reproduce the occupied electronic
manifold. More precisely, our purpose was to count how many occupied
WFs are centered at the different Ni cations, and how many at the
surrounding oxygen anions, and to characterize them. Further, we
wanted to run our analysis without having to make any assumption on
the precise character of the occupied Ni and O orbitals, which
complicated the choice of the seed functions that are needed for an
efficient maximal-localization calculation. Nevertheless, we found the
following robust strategy to proceed.  We considered the whole
occupied manifold and sought to extract from it ({\em i.e.}, to
disentangle) a set of 2$\times$(144+10) WF functions, where
2$\times$144 = 2$\times$3$\times$48 is the total number of O-$2p$
orbitals available in our 80-atom supercell and 2$\times$10 =
2$\times$5$\times$2 is the number of Ni-$3d$ orbitals corresponding to
two specific Ni atoms. (Note that we ran separate WF optimizations for
the spin-up and spin-down channels.) Hence, for our initial WF seeds,
we used 3 generic $p$ orbitals centered at each of the O anions in our
cell, and 5 generic $d$ orbitals centered at two neighboring Ni
cations; this couple of Ni cations were chosen to be first-nearest
neighbors, so that we considered one Ni$_{\rm L}$ and one Ni$_{\rm
  S}$. The basic qualitative results of this optimization were the
same for all the nickelates considered, and thus the following
discussion is not compound specific.

Our optimization renders 2$\times$3 WFs centered at each oxygen anion
({\em i.e.}, 3 spin-up WFs and 3 -- very similar -- spin-down WFs),
suggesting that all oxygens in our nickelates are in a $2-$ oxidation
state. The oxygen-centered WFs have a clear $p$ character, as can be
appreciated in Figures~\ref{fig4}.c and d. We also obtained 2$\times$3
$t_{2g}$-like WFs centered at each of the two considered Ni atoms,
indicating that the $t_{2g}$ states are fully occupied and there is no
magnetic moment associated to them. Further, we obtained 2
$e_{g}$-like spin-up WFs centered at the Ni$_{\rm L}$ site (see
Figures~\ref{fig4}.a and b), indicating that this cation is in a $2+$
oxydation state and has a significant magnetic moment associated to
it. Finally, as regards the other seed functions centered at the
chosen Ni$_{\rm L}$ (2 spin-down $d$ orbitals) and Ni$_{\rm S}$
(2$\times$2 $d$ orbitals) atoms, they did not lead to any WF centered
at those sites. Instead, the maximal-localization procedure resulted
in WFs centered at Ni and R cations in the vicinity of the considered
Ni$_{\rm L}$--Ni$_{\rm S}$ pair. Thus, in particular, it was
impossible to localize any $e_{g}$-like WFs at a Ni$_{\rm S}$ site,
which strongly indicates that these Ni cations are in a $4+$ oxidation
state. These conclusions were ratified by considering larger clusters
of Ni sites for the WF optimization, as well as individual Ni's and/or
optimizations in which the oxygen bands were not included.

Hence, the Wannier analysis yields a picture of strong charge
disproportionation between the Ni sites, which is clearly at odds with
the quantitatively similar behavior discussed in the section above. To
resolve this apparent contradiction, we need to inspect in more detail
the obtained oxygen-centered WFs.

Figures~\ref{fig4}.c and d report representative results, for
SmNiO$_{3}$, of the O-$2p$-like Wannier functions oriented along the
Ni--O--Ni bonds. As it is clearly visible in the figures for both spin
channels, these WFs have their centers significantly shifted towards
the Ni$_{\rm S}$ cations and away from the Ni$_{\rm L}$ sites. The
shift -- as quantified by the distance between the oxygen position and
WF center -- is $\sim 0.231$~\AA~in average. Hence, while the Ni$_{\rm
  S}$ cations appear to be in a $4+$ state when we count how many WFs
are centered at them, they also {\em receive} a significant fraction
of electrons coming from the surrounding oxygens, with which they are
strongly hybridized. Hence, this is the explanation why quantititative
measures of the charge around the Ni$_{\rm S}$ ions renders results
that are similar to those of the Ni$_{\rm L}$ and suggest valence
state much more reduced than the expected $4+$. Across the series, we
observe that O-$2p$ centers get closer to the Ni$_{\rm S}$ cations for
smaller rare earths, increasing the localization of the charge on
Ni$_{\rm S}$. Additionally, the O-$2p$ WFs behave similarly
irrespective of their spin polarization, so that their shifting does
not result in any magnetic moment at the Ni$_{\rm S}$ sites.

The Wannier analysis therefore leads to the conclusion that a full
disproportionation occurs in the system, with clearly distinct
Ni$_{\rm S}^{4+}$ (low-spin, non-magnetic) and Ni$_{\rm L}^{2+}$
(high-spin, magnetic) sites. Simultaneously, the O-$2p$ WFs approach
the Ni$_{\rm S}^{4+}$ sites, ultimately yielding Ni$_{\rm S}$ and
Ni$_{\rm L}$ that are nearly equivalent from the point of view of the
charge (static and dynamics) around them.

\section*{Discussion}

Our results thus appear to be compatible with the disproportionation
effects originally proposed to occur at the
MIT~\cite{alonso1999metal,PRB-Medarde-PNO,PRB-medarde-2009}. While for
a long time the nickelates were believed to possess an orthorhombic
$Pbnm$ symmetry in the insulating
phase~\cite{Review-medarde1997structural}, it is now established that
they adopt a monoclinic $P2_1/n$ phase at low
temperatures~\cite{alonso1999metal,PRB-Medarde-PNO}. This phase
exhibits a breathing distortion whose magnitude decreases with
increasing the tolerance factor of the
perovskite~\cite{PRB-Medarde-PNO}, concomitantly accompanied by a
charge disproportionation $\delta$ between the two Ni sites, leading
to a Ni$_{\rm S}^{(3+\delta)+}$+Ni$_{\rm L}^{(3-\delta)+}$
configuration~\cite{PRB-medarde-2009}. Our results are reminiscent of
this picture, with the observation of a subsequent breathing
distortion modulated by the rare earth and a charge disproportionation
between Ni sites.

Nevertheless, they are also compatible with the model of Mizokawa {\em
  et al}~\cite{Sawatzky-covalence}, as well as with recent Dynamical
Mean Field Theory (DMFT)
studies~\cite{park2012site,AGeorges-lowenergyNickelates,NNO-SNO-Optical2015,PRL110-126404,PRB89245133},
proposing a ligand-hole structure in rare-earth nickelates. Indeed,
our Wannier analysis indicates that we have a $3d^8$ electronic
configuration for the Ni$_{\rm L}$ cations. More importantly, it is
also compatible with the $3d^8{\underline L}^2$ configuration proposed
for the Ni$_{\rm S}$ sites. More precisely, the notation ${\underline
  L}^2$ stands for the two oxygen holes that are shared by the oxygens
in the O$_{6}$ octahedron surrounding a Ni$_{\rm S}$ cation. According
to our Wannier analysis, such a situation would correspond to having
all six oxygens around Ni$_{\rm S}$ sharing the 2$p$ electrons that
occupy orbitals along the Ni--O--Ni bonds. Further, given that our
integrated and dynamical Born charges suggest that the Ni$_{\rm S}$
and Ni$_{\rm L}$ sites host a similar number of electrons, our results
do in fact point to a situation in which each O$_{6}$ cage shares
approximately two electrons with the Ni$_{\rm S}$ at its center,
exactly as in the ligand-hole picture.

As regards the magnetic moments at the different Ni sites, our results
are also clear and compatible with both proposed pictures: Ni$_{\rm
  L}$ bears a magnetic moment approaching 2 $\mu_B$, as it corresponds
to having Ni$^{2+}$ in a high-spin configuration. Then, Ni$_{\rm S}$
has no magnetic moment associated to it, as it would correspond to a
nominal Ni$^{4+}$ low-spin configuration. The latter result is partly
a consequence of the fact that the oxygen--Ni$_{\rm S}$ shared
electrons are spin paired, which can be interpreted as a ligand-hole
screening.

\begin{figure}[h!]
\begin{center}
\resizebox{8.6cm}{!}{\includegraphics{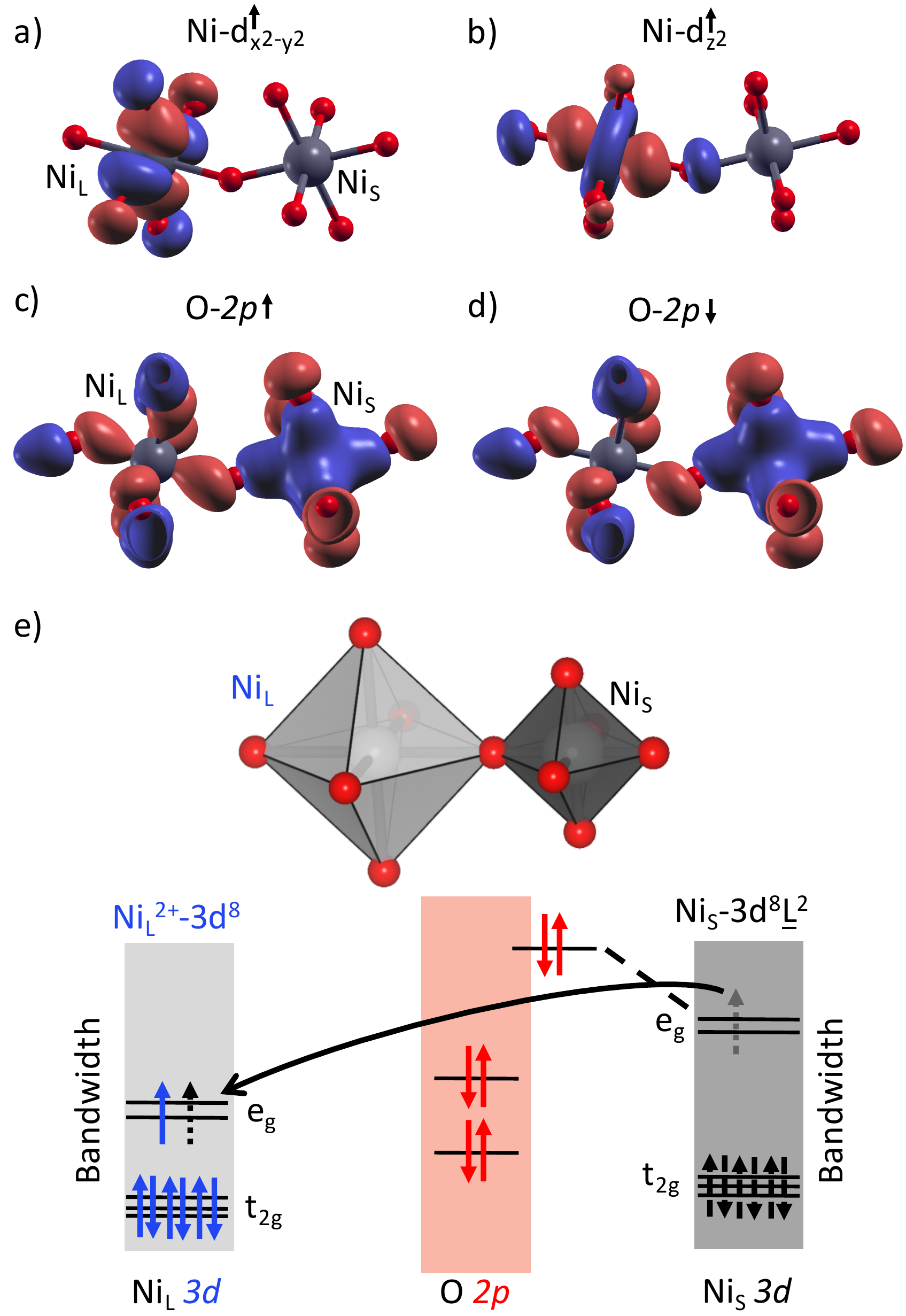}}
\end{center}
\caption{{\bf Orbital occupancies in rare-earth nickelates.}  (a)
  $d_{x^2-y^2}^{\uparrow}$ maxi localized Wannier function on two
  consecutive Ni sites. The Ni$_{\rm L}$ site bears a spin up. (b)
  $d_{z^2}^{\uparrow}$ maxi localized Wannier function on two
  consecutive Ni sites. The Ni$_{\rm L}$ site bears a spin up. (c)
  O-$2p$ maxi localized Wannier functions oriented along the Ni-O
  bonds on the two consecutive Ni sites in the spin up channel.  (d)
  O-$2p$ maxi localized Wannier functions oriented along the Ni-O
  bonds on the two consecutive Ni sites in the spin down channel. (e)
  Schematic picture of the electronic structure of rare-earth
  nickelates ground state.}
\label{fig4}
\end{figure}

Our first-principles calculations therefore reconcile the two visions
proposed to occur in the ground state of rare-earth nickelates. The
electronic structure is summarized in Figure~\ref{fig4}.e and is based
on having i) a full charge disproportionation between Ni$_{\rm L}$ and
Ni$_{\rm S}$ sites accompanied by a breathing mode and ii) 2 electrons
from surrounding oxygens shared with the depleted Ni$_{\rm S}$ site,
leaving the impression of two similar Ni$^{2+}$ sites in the
system. Finally, this electronic structure lifts the orbital
degeneracy appearing in the high temperature phase and is compatible
with the small Jahn-Teller distortion observed in our computed ground
states.

Although the electronic structure is similar between all nickelates,
the rare-earth cation does have an impact on the level of covalency of
the system through their induced lattice distortions. An example is
their influence on the magnetic moment of the Ni$_{\rm L}$
cations. Note that our computed Ni$_{\rm L}$ magnetic moments are
always far from the nominal value of 2$\mu_B$, as a consequence of the
the hybridization between Ni-$3d$ and O-$2p$ states. This can be seen
in Figures~\ref{fig4}.c and d where slightly larger overlaps between
O-$2p$ orbitals over the Ni$_{\rm L}$ site arise in the spin up
channel (this Ni$_{\rm L}$ site has a net spin up). From
Figure~\ref{fig3}.e, the level of covalency clearly increases with the
tolerance factor (a ferromagnetic calculation yields similar
conclusion, see Supplementary Material). This is further confirmed by
the increase of the average BECs on both Ni$_{\rm S}$ and Ni$_{\rm L}$
following their interprettion of such effects in other perovskite
oxides~\cite{ghosez1995born}. We also observe increasing O-$2p$
overlaps with Ni$_{\rm L}$ sites going from R=Y to Pr.

The level of covalency seems to correlate with the stability of the
magnetic ordering and the insulating phase (see Table~\ref{tab1}). On
one hand, with increasing covalency, the energy difference between the
AFM (S or T) and FM solutions increases strongly. On the other hand,
the insulating gap decreases. In order to further corroborate these
trends, we performed additional calculations on SmNiO$_3$ by applying
a hydrostatic pressure of $\pm$8\% on the ground state volume and
relaxed the atomic positions for each magnetic ordering. Under
compression, we find a sizable enhancement of the covalent character
as $\mu_{\text Ni_{\text L}}$ decreases to 1.103~$\mu_B$. The
stability of the S-AFM ordering with respect to the FM solution is
therefore doubled ($\Delta E$=-293 meV per 80-atom cell). Eventually,
the band gap is decreased by 0.11~eV with respect to the ground state
value. Under expansion, we observe a weakening of the covalent
character of the system with $\mu_{\text Ni_{\text L}}$ increasing to
1.308~$\mu_B$; simultaneously, the stability of the complex AFM
ordering is roughly reduced by a factor of 2 ($\Delta$E =-67 meV per
80-atom cell).

\begin{figure}[h!]
\begin{center}
\resizebox{8.6cm}{!}{\includegraphics{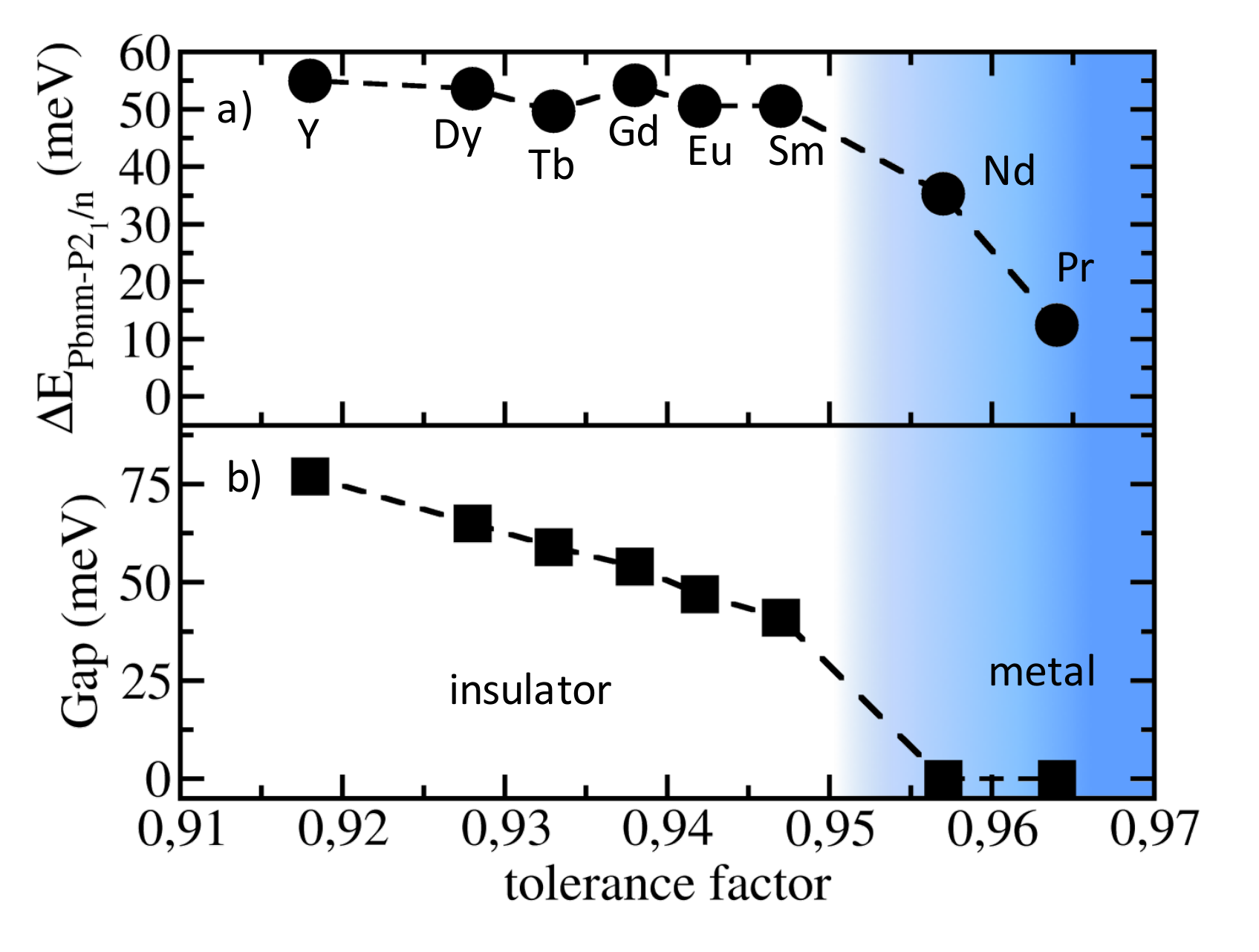}}
\end{center}
\caption{ {\bf Metal-insulator phase transition} (a) Energy difference
  (in meV) per 20 atoms unit cell between the relaxed monoclinic and
  orthorhombic phases for a ferromagnetic solution; (b) Band gap (in
  meV) of the ferromagnetic monoclinic phase.}
\label{fig5}
\end{figure}

Finally, the rare-earth atom is known to play key role in the nature
of the MIT. Experimentally it is observed that T$_{\text {MI}}$ is
different from the magnetic-ordering transition temperature T$_{\text
  N}$ for all nickelates except for those in which the rare earth is
Pr or Nd. Interestingly, our calculations reflect this differentiated
behavior. As already mentioned, we obtain an insulating solution for
the AFM monoclinic ground state of all considered nickelates. Then,
when we consider the $P2_1/n$ structure with a ferromagnetic spin
arrangement, we also obtain an insulating phase for all R cations
ranging from Y to Sm; the corresponding band gaps range from 77 to
41~meV and we observe a relatively large energy gain with respect to
the orthorhombic phase (see Figures~\ref{fig5}.a and b). Thus, the
breathing distortion and disproportionation effects seem sufficient to
open the band gap in these compounds, irrespective of the spin
arrangement. As a consequence, our results indicate that these
nickelates can potentially present an insulating, spin-disordered
phase, as they indeed do experimentally. In contrast, for the
monoclinic phase of NdNiO$_3$ and PrNiO$_3$, the FM spin configuration
is found to be metallic; further, the stability of the low-symmetry
strucure with respect to the orthorhombic one drastically
decreases. The complex antiferromagnetic ordering therefore appears to
be a necessary condition for the MIT to occur in these two compounds.

Considering the case of SmNiO$_3$ under 8\% of compression, we unveil
a similar behaviour to bulk PrNiO$_3$ and NdNiO$_3$, since the level
of covalency increases. These results suggest that it is possible to
control of the electronic and magnetic structures of these compounds
by tuning the level of covalency in the system. We also emphasize that
these observations again support our choice of a small Hubbard U
correction on Ni-$3d$ sites and demonstrate that DFT+U methods can
capture the key physical properties of correlated systems.

In summary, we have used first-principles methods to investigate the
ground state electronic structure of rare-earth nickelates. Our DFT
simulations using a small Hubbard correction on Ni-$3d$ states
reproduce all features reported from experiments (insulating
character, disproportionation effects, covalency, complex
antiferromagnetic structures, structural trends). In particular, we
show that the insulating phase is characterized by a clear-cut split
of the electronic states of the two Ni sites, which can be strictly
described as being low-spin $4+$ and high-spin $2+$. At the same time,
our simulations reveal a shift of the oxygen-$p$ orbitals toward the
depleted Ni cations, so that, ultimately, from the point of view of
the integrated charge, the two Ni sites appear to be nearly
identical. These findings are clearly reminiscent of the various
pictures proposed in the literature to explain the ground state of
these compounds, which can thus be reconciled according to our
results. Finally, we unveil that a control of the level of covalency
between oxygens and transition metal ions provides an alternative
pathway to tune the electronic and magnetic phases in late
transition-metal oxide perovskites.

\section*{Methods}
First-principles calculations were performed with the Vienna Ab-initio
Simulation Package (VASP) package~\cite{VASP1,VASP2}. Magnetism was
treated only at the collinear level. We used a 3$\times$6$\times$2
$\Gamma$-centered K-point mesh. The cut-off was set to 500 eV. We used
PAW pseudopotentials~\cite{PAW} with the following electron
configurations: 4s$^2$3d$^8$ (Ni), 2s$^2$2p$^4$ (O),
4s$^2$4p$^6$5s$^2$4d$^1$ (Y), 4s$^2$4p$^6$5s$^2$4f$^1$ (Pr, Nd, Sm),
4p$^6$5s$^2$4f$^1$ (Gd, Eu, Tb, Dy). We did not treat explicitly the
$f$ electrons as they order at very low
temperature~\cite{MagHoNiO3-possibleNonCol} and they were included in
the pseudopotential. Full geometry relaxations were performed until
forces were lower than 0.001 eV/\AA~ and energy was converged to
1$\times$10$^{-7}$ eV. Born Effective Charges were computed using
density functional perturbation theory~\cite{RMP-Baroni}. Symmetry
adapted modes allowing the extraction of lattice distortion amplitudes
were performed using the Bilbao crystallographic
server~\cite{Amplimodes1,Amplimodes2}. The results presented in
Figures~\ref{fig2}.d, e and f have been obtained by freezing single
lattice distortion in a cubic reference that has the same lattice
parameters for all reported compounds. This allows to extract
quantities that are strain independent for all nickelates. Wannier
functions reported in Figures~\ref{fig4}.a and b are plotted for
isosurfaces equal to 2. Wannier functions reported in
Figures~\ref{fig4}.c and d are plotted for isosurfaces equal to 9.

\section*{acknowledgments}
 Work supported by the ERC grant MINT (contract \#615759) and by
 National Research Fund, Luxembourg through a Pearl Grant
 (FNR/P12/4853155). J. Varignon acknowledges Ph. Ghosez and A. Mercy
 for fruitful discussions.




\begin{thebibliography}{60}
\bibitem{noteOrdreMag} We have also considered usual A, C and G-type
  AFM orderings but these solutions are always found higher in energy
  than the FM state by at least 400 meVs.
\bibitem{gap0eV} The gap is eventually reduced to 91 meV without
  Coulombic correction on the Ni $3d$ states.
\end{thebibliography}
\end{document}